\documentclass[10pt]{IEEEtran}
\usepackage{color}
\usepackage{times}
\usepackage{amssymb}
\usepackage{amsmath}
\usepackage{epsfig}
\usepackage{cite}
\usepackage{verbatim}
\usepackage{enumerate}
\usepackage{subfigure}
\usepackage{multicol}
\usepackage{pstricks}
\usepackage{pst-node}
\usepackage{amsmath}
\usepackage{graphicx}
\usepackage{setspace}
\usepackage{algpseudocode}
\usepackage{algorithm}
\usepackage{cases}

\usepackage{picinpar}

\begin{document}
\title{ Approaching the Capacity of Wireless Networks through Distributed Interference Alignment}
\author{\authorblockN{Krishna Gomadam, Viveck R. Cadambe, Syed A. Jafar}\\
\authorblockA{Electrical Engineering and Computer Science\\
University of California, Irvine, CA 92697-2625\\ Email: \{kgomadam, vcadambe,
syed\}@uci.edu}} \maketitle \IEEEpeerreviewmaketitle

\begin{abstract}
Recent results establish the optimality of  interference alignment to approach the Shannon capacity of interference networks at high SNR. However, the extent to which interference can be aligned over a finite number of signalling dimensions remains unknown. Another important concern for interference alignment schemes is the requirement of global channel knowledge. In this work we provide examples of iterative algorithms that utilize the reciprocity of wireless networks to achieve interference alignment with only local channel knowledge at each node. These algorithms also provide numerical insights into the feasibility of interference alignment that are not yet available in theory.
\end{abstract}

\section{Introduction}
The recent emergence of the idea of interference alignment for
wireless networks has shown that the capacity of wireless networks
can be much higher than previously believed \cite{Cadambe_Jafar_int}. The canonical example of
interference alignment is a communication scenario where, regardless
of the number of interferers, every user is able to access one half
of the spectrum free from interference from other users
\cite{Cadambe_Jafar_int}. For the 
interference channel with $K$ transmitters and $K$ receivers and random, time
varying channel coefficients drawn from a continuous distribution, reference \cite{Cadambe_Jafar_int}
characterizes the network sum capacity as
\begin{equation}
C_\Sigma(SNR) = \frac{K}{2} \log(SNR) + o(\log(SNR))\label{eq:dofint}
\end{equation}
 so that the capacity per user
is $\frac{1}{2}\log(SNR)+o(\log(SNR))$. Here SNR is defined as the total
transmit power of all the transmitters in the network when the local
noise power at each node is normalized to unity. The $o(\log(SNR))$ term, by definition, becomes negligible compared to $\log(SNR)$ at high SNR. Therefore the accuracy of the capacity approximation in (\ref{eq:dofint}) approaches $100\%$ at high SNR. Since the capacity
of a single user in the absence of all interference is
$\log(SNR)+o(\log(SNR))$, the main
result of \cite{Cadambe_Jafar_int} may be summarized as:

``\emph{At high SNR, every user in a wireless interference network is
(simultaneously and almost surely) able to achieve (nearly) one half of the capacity that he could achieve in the absence of all interference.}"

The capacity-optimal achievable scheme within $o(\log(\mbox{SNR}))$ is shown to be the interference alignment scheme. Interference alignment on the $K$ user interference channel refers to the idea of constructing signals in such a way that they cast overlapping shadows over one half of the signal space observed by each receiver where they constitute interference, leaving the other half of the signal space free of interference for the desired signal. This approach reveals the sub-optimality of the cake-cutting view of spectrum allocation  between co-existing wireless systems because, essentially, everyone gets ``half the cake''.

Interference alignment schemes are presented in \cite{Cadambe_Jafar_int}  in the form of closed form expressions for the transmit precoding matrices. However, these closed form expressions require global channel knowledge which can be an overwhelming overhead in practice. Moreover, closed form solutions have only been found in certain cases. In general, analytical solutions to interference alignment problem are difficult to obtain and even the feasibility of interference alignment over a limited number of signalling dimensions is an open problem. In this paper we explore distributed interference alignment algorithms to accomplish the following objectives.
\begin{itemize}
\item Require only local channel knowledge at each node. Specifically, each receiver is assumed to know only the channel to its desired transmitter and the covariance matrix of its effective noise (consisting of the AWGN and the interference from all other users).
\item Provide numerical insights into the feasibility of alignment.
\end{itemize}

We propose iterative algorithms that take a cognitive approach to interference management and utilize only the local side information available naturally due to the reciprocity of wireless networks. The two key properties can be summarized as follows.
\begin{itemize}
\item{\it Cognitive Principle:} Unlike selfish approaches studied in prior work where each transmitter tries to maximize his own rate by transmitting along those signaling dimensions where his desired receiver sees the least interference, we follow an unselfish approach where each transmitter primarily tries to minimize the interference to unintended receivers. Since avoiding interference to unintended receivers is the defining feature of cognitive radio \cite{cogtutorial} the unselfish approach is a cognitive approach. The cognitive approach is found to lead to interference alignment, and is thus capable of approaching network capacity at high SNR.
\item{\it Reciprocity:} For a given transmitter, learning how much interference is caused at unintended receivers can require too much side information, and is one of the key challenges for cognitive radio systems. However, this information is naturally available because of the reciprocity of the channel for networks where two-way communication is based on time-division duplex operation with synchronized time-slots. Due to reciprocity, the signalling dimensions along which a receiving node sees the least interference from other users  are also the same signalling dimensions along which this node will cause the least interference to other nodes in the reciprocal network where all transmitters and receivers switch roles.
\end{itemize}

The paper is organized as follows. In the next section, we review some optimization approaches for interference networks in existing literature. In Section \ref{sec:recalign} we state the open problem of determining the feasibility of interference alignment over a limited number of signaling dimensions. The same section also describes the reciprocity property of interference alignment. In  Section \ref{sec: algo} we present iterative interference alignment algorithms. In Section \ref{sec: results}, we show that these algorithm can achieve performance close to the theoretical results and discuss a few applications. We conclude with Section \ref{sec: concl}.

{\it Notation:} We use lower case for scalars, upper case for vectors and bold font to denote matrices. ${\bf A}_{\star d}$ represents the $d^{th}$ column of matrix ${\bf A}$. ${\bf I}_{d}$ represents the $d\times d$ identity matrix. Tr$[{\bf A}]$ denotes the trace of the matrix ${\bf A}$ and ${\bf A}^\dagger$ is the conjugate transpose of matrix ${\bf A}$. Finally, $\mathcal{K}\triangleq\{1,2,\cdots, K\}$ is the index set of $K$ users.

\section{Interference Optimization Approaches}
The optimality of interference alignment schemes at high SNR is interesting because these schemes treat all interference as noise and require no multi-user detection. Achievable schemes based on treating interference as noise have been explored extensively over the last decade. Prominent among these are the interference avoidance and iterative waterfilling algorithms where each transmitter acts selfishly to align its transmissions along those directions where its desired receiver sees the least interference \cite{rose_ia,
rose_interfering, game_fair, ia_game, iterative_ulukus_yates}, and network duality approaches \cite{wc_book_tse, transmit_beamforming_farrokhi, network_dual_rao,babadi-2007} that are based on the reciprocity of the wireless propagation channel. 
\subsection{Interference Avoidance and Iterative Waterfilling}
Iterative algorithms are commonly used for various resource allocation problems, such as "interference avoidance" and "iterative waterfilling". However, the philosophy of interference alignment is quite distinct from both iterative waterfilling and interference avoidance. With iterative waterfilling/interference avoidance algorithms \cite{yu_iterative_waterfilling, rose_ia, etkin_tse_2007}, each transmitter tries to do what is best for his own receiver, i.e., each transmitter allocates its power in a manner best suited for his desired receiver. With interference alignment each transmitter tries to minimize the interference he causes to other receivers. The interference alignment schemes in \cite{Jafar_Shamai, Cadambe_Jafar_int, Cadambe_Jafar_X, Cadambe_Jafar_XFB} show that for interference networks, the ``do no harm'' approach is much more powerful, and is in fact capacity-optimal within $o(\log(\mbox{SNR}))$, than the ``help yourself'' approach  of interference avoidance and iterative waterfilling schemes.

\begin{figure}
\center
\input{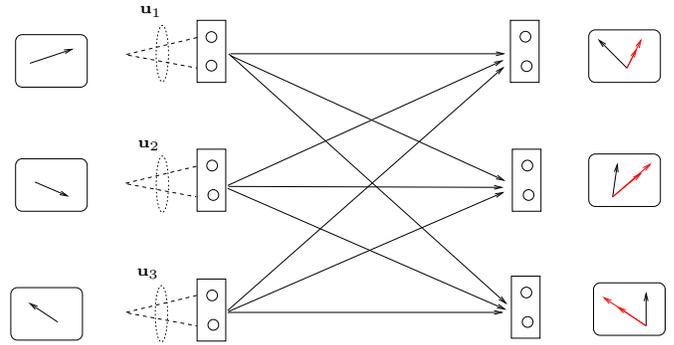}
\caption{Interference alignment solution for the three user two
antenna case. The arrows in the red indicate the direction of the
interference.}\label{fig: nash}
\end{figure}

From a game-theoretic perspective, interference avoidance and iterative waterfilling algorithms lead to a stable operating point commonly known as the Nash equilibrium. At Nash equilibrium, there is no incentive for any user to unilaterally change his transmit strategy. Often these points are not optimum from a network perspective and indicate inefficiency in wireless network operation \cite{comp_coll_miso_int, leshem-2007}. Interestingly, interference alignment is not a Nash equilibrium point if the goal of each user is to maximize his own rate. Fig. \ref{fig: nash} shows  the interference alignment solution for the three user two antenna case. Notice that interfering signals are co-linear at each receiver while the desired signal may not be exactly orthogonal to the interference - a price paid for interference alignment. It can be easily observed that the interference alignment solution is not a Nash equilibrium point. Fixing the transmit strategy for users 2 and 3, the best strategy for user 1 is to choose ${\bf u}_1$ such that his signal is orthogonal to the interference at receiver 1. Although this strategy is good for user 1, it will destroy interference alignment at receivers 2 and 3. Thus Fig. \ref{fig: nash} clearly highlights the difference between the optimal strategy of interference alignment  and the selfish strategy of iterative waterfilling or interference avoidance. 

Iterative schemes have also been used to implicitly achieve interference alignment on the $2$ user $X$ channel in \cite{MMK,MMKreport1,MMKreport2}. However, for the $2$ user $X$ channel interference alignment can be explicitly achieved with roughly the same amount of channel knowledge as required by the iterative schemes, without the need for an iterative process \cite{Jafar_Shamai}. The iterative schemes of \cite{MMK} are specialized for the $2$ user $X$ channel and generalizations to $X$ networks and interference networks with more than $2$ users are not straightforward.

\subsection{Network Duality}
Another approach taken in prior work is to exploit the duality relationships enabled by the reciprocity of the propagation channel. For example, network duality ensures that the same set of signal to interference and noise ratios (SINRs) can be achieved in the original and the reciprocal network with the same total transmit power \cite{wc_book_tse, transmit_beamforming_farrokhi}. Network duality is used in \cite{transmit_beamforming_farrokhi, network_dual_rao} to minimize the total transmit power required to support a feasible rate vector. Reciprocity of propagation channels is used in \cite{babadi-2007} for optimal frequency allocation problem. 

In this work we provide examples of iterative algorithms to achieve interference alignment on wireless interference channels. These algorithms combine elements of all the above-mentioned approaches, especially \cite{MMK} and \cite{babadi-2007}.

\section{System Model}
Consider the $K$-user MIMO interference channel where the $k^{th}$ transmitter and receiver are equipped with $M^{[k]}$ and $N^{[k]}$ antennas respectively. Note that the antennas could represent symbol extensions in time or frequency as well. However, if the antennas correspond to symbol extensions over orthogonal dimensions (time, frequency slots) then the channel matrices will have a diagonal structure. The channel is defined as:
\begin{eqnarray*}
Y^{[k]}(n)=\sum_{l=1}^K {\bf H}^{[kl]}(n)X^{[l]}(n)+Z^{[k]}(n), ~~\forall k\in\mathcal{K}
\end{eqnarray*}
where, at the $n^{th}$ channel use, $Y^{[k]}(n), Z^{[k]}(n)$ are the $N^{[k]}\times 1$ received signal vector and the zero mean unit variance circularly symmetric additive white Gaussian noise vector (AWGN) at receiver $k$, $X^{[l]}(n)$ is the $M^{[l]}\times 1$ signal vector transmitted by transmitter $l$, and ${\bf H}^{[kl]}(n)$ is the $N^{[k]}\times M^{[l]}$ matrix of channel coefficients between transmitter $l$ and receiver $k$. The transmit power at transmitter $l$ is $\mbox{E}[||X^{[l]}||^2]=P^{[l]}$. 

For the $K$ user interference channel defined above, we also define a reciprocal channel, where the role of transmitters and receivers are switched. For every variable on the original channel, the corresponding variable on the reciprocal channel is denoted with a left arrow on top. The reciprocal channel is defined as:
\begin{eqnarray*}
\overleftarrow{Y}^{[k]}(n)=\sum_{l=1}^K \overleftarrow{\bf H}^{[kl]}(n)\overleftarrow{X}^{[l]}(n)+\overleftarrow{Z}^{[k]}(n), ~~\forall k\in\mathcal{K}
\end{eqnarray*}
where, at the $n^{th}$ channel use, $\overleftarrow{Y}^{[k]}(n), \overleftarrow{Z}^{[k]}(n)$ are the $M^{[k]}\times 1$ received signal vector and the zero mean unit variance circularly symmetric additive white Gaussian noise vector (AWGN) at receiver $k$ which is equipped with $M^{[k]}$ antennas, $\overleftarrow{X}^{[l]}(n)$ is the $N^{[l]}\times 1$ signal vector transmitted by transmitter $l$, and $\overleftarrow{\bf H}^{[kl]}(n)={\bf H}^{[lk]\dagger}(n)$ is the $M^{[k]}\times N^{[l]}$ matrix of channel coefficients between transmitter $l$ and receiver $k$. The transmit powers at transmitter $l$ on the reciprocal channel is $\mbox{E}[||\overleftarrow{X}^{[l]}||^2=\overleftarrow{P}^{[l]}$. The channel use index $n$ is henceforth suppressed to avoid cumbersome notation.

\section{Interference Alignment over Limited Dimensions - An Open Problem}\label{sec:recalign}
References \cite{Cadambe_Jafar_int} and \cite{Cadambe_Jafar_X} present interference alignment schemes that are constructed over symbol extensions of time-varying channels. It is shown that by using \emph{long} symbol extensions the degrees of freedom achieved per dimension approach arbitrarily close to the theoretical outerbound, thereby establishing the degrees of freedom of time-varying interference and $X$ networks. However, the extent to which interference can be aligned over a \emph{limited} number of dimensions remains an open problem. As a consequence, the maximum number of degrees of freedom that can be achieved through  alignment of interference signal vectors is not known in general. Note that interference alignment can also be accomplished in terms of signal \emph{levels} rather than signal vectors by using structured codes (multilevel and lattice codes) as shown in \cite{Bresler_Parekh_Tse, Cadambe_Jafar_Shamai}. However, in this work our focus is on interference alignment through signal vectors. We first review the interference alignment problem and the reciprocity property of interference alignment.


Let $d^{[k]}\leq\min(M^{[k]},N^{[k]}), k\in\mathcal{K}$ denote the degrees of freedom for user $k$'s message. 

{\it Precoding at Transmitter:} Let ${\bf V}^{[k]}$ be an $M^{[k]}\times d^{[k]}$ matrix whose columns are the orthonormal basis of the transmitted signal space of user $k$. Mathematically, the transmitted signal vector of user $k$ is given by:
\begin{eqnarray}
{X}^{[k]}=\sum_{d=1}^{d^{[k]}}{\bf V}^{[k]}_{[\star d]}\overline{X}^{[k]}_d={\bf V}^{[k]}{\overline{X}}^{[k]}, ~~\overline{X}^{[k]}\sim\mathcal{N}\left(0,\frac{P^{[k]}}{d^{[k]}}{\bf I}_{d^{[k]}}\right)
\end{eqnarray}
Each element of the $d^{[k]}\times 1$ vector ${\overline{X}}^{[k]}$ represents an independently encoded Gaussian codebook symbol with power $\frac{P^{[k]}}{d^{[k]}}$ that is beamformed with the corresponding vector of ${\bf V}^{[k]}$. 

{\it Remark:} For interference alignment or the achievability of the degrees of freedom it suffices if the beamforming vectors are linearly independent. However, we assume the beamforming vectors are orthonormal in the formulation above. Note that this does not affect the feasibility of interference alignment, and it naturally leads to the iterative algorithm to be presented in this paper.

{\it Interference Suppression at Receiver:} Let ${\bf U}^{[k]}$ be an $N^{[k]}\times d^{[k]}$ matrix whose columns are the orthonormal basis of the interference-free desired signal subspace at receiver $k$. The $k^{th}$ receiver filters its received signal to obtain:
\begin{eqnarray}
\overline{Y}^{[k]}={\bf U}^{[k]\dagger} Y^{[k]}
\end{eqnarray}
If interference is aligned into the null space of ${\bf U}^{[k]}$ then the following condition must be satisfied:
\begin{eqnarray}
{\bf U}^{[k]\dagger}{\bf H}^{[kj]}{\bf V}^{[j]} &=& 0, \forall j\neq k\\
\mbox{rank}\left({\bf U}^{[k]\dagger}{\bf H}^{[kk]}{\bf V}^{[k]}\right)&=&d_k
\end{eqnarray}
In other words the desired signals are received through a $d^{[k]}\times d^{[k]}$ full rank channel matrix $$\overline{\bf H}^{[kk]}\triangleq {\bf U}^{[k]\dagger}{\bf H}^{[kk]}{\bf V}^{[k]} $$ while the interference is completely eliminated. The effective channel for user $k$ is then expressed as:
\begin{eqnarray}
\overline{Y}^{[k]}&=&\overline{\bf H}^{[kk]}{\overline{X}}^{[k]}+\overline{Z}^{[k]}
\end{eqnarray}
where $\overline{Z}^{[k]} \sim\mathcal{N}\left(0,I\right)$ is the effective $d^{[k]}\times 1$ AWGN vector at receiver $k$. The rate achieved on this channel is:
\begin{eqnarray}
R^{[k]}&=&\log\left|{\bf I}_{d^{[k]}}+\frac{P^{[k]}}{d^{[k]}}\overline{\bf H}^{[kk]}\overline{\bf H}^{[kk]\dagger}\right|\\&=&d^{[k]}\log(P^{[k]})+o(\log(P^{[k]}))
\end{eqnarray}
Thus, $d^{[k]}$ degrees of freedom are achieved by user $k$.

\subsection{Feasibility of Alignment}
Given the channel matrices ${\bf H}^{[kj]}, k,j\in\mathcal{K}$, we say that the degrees of freedom allocation $(d^{[1]}, d^{[2]}, \cdots, d^{[K]})$ is feasible if there exist transmit precoding matrices ${\bf V}^{[k]}$ and receive interference suppression matrices ${\bf U}^{[k]}$:
\begin{eqnarray}
{\bf V}^{[k]}:M^{[k]}\times d^{[k]},&&{\bf V}^{[k]\dagger}{\bf V}^{[k]}={\bf I}_{d^{[k]}}\\
{\bf U}^{[k]}:N^{[k]}\times d^{[k]},&&{\bf U}^{[k]\dagger}{\bf U}^{[k]}={\bf I}_{d^{[k]}}
\end{eqnarray}
such that
\begin{eqnarray}
{\bf U}^{[k]\dagger}{\bf H}^{[kj]}{\bf V}^{[j]} &=& 0, \forall j\neq k\label{eq:cross}\\
\mbox{rank}\left({\bf U}^{[k]\dagger}{\bf H}^{[kk]}{\bf V}^{[k]}\right)&=&d_k, ~\forall k\in\mathcal{K} \label{eq:direct}
\end{eqnarray}

{\it Remark:} Suppose all the elements of the channel matrices are randomly and independently generated from continuous distributions and ${\bf V}^{[k]},{\bf U}^{[k]}, k\in\mathcal{K}$ can be found to satisfy condition (\ref{eq:cross}). Then, condition (\ref{eq:direct}) will also be satisfied with probability 1. This is because the direct channel matrices ${\bf H}^{[kk]}$ do not appear in condition (\ref{eq:cross}). So the choice of transmit and receive filters ${\bf V}^{[k]},{\bf U}^{[k]}, k\in\mathcal{K}$ to satisfy (\ref{eq:cross}) does not depend on the direct channel matrices ${\bf H}^{[kk]}$. Since ${\bf H}^{[kk]}$ is independent of ${\bf V}^{[k]},{\bf U}^{[k]}$ and all its elements are randomly generated from a continuous distribution (i.e. it lacks any special structure), the product matrix ${\bf U}^{[k]T}{\bf H}^{[kk]}{\bf V}^{[k]}$ has full rank with probability $1$. Thus, for random MIMO channels without time-extensions, if (\ref{eq:cross}) can be satisfied then (\ref{eq:direct}) is automatically satisfied almost surely as well. However, if time-extensions are considered then the channel matrices may have a block diagonal structure and (\ref{eq:direct}) cannot be taken for granted.

The solution to the feasibility problem is not known in general. In other words, given a  set of randomly generated channel matrices and a degree-of-freedom allocation $(d^{[1]}, d^{[2]}, \cdots, d^{[K]})$, it is not known if one can almost surely find transmit and receive filters that will satisfy the feasibility conditions. The distributed interference algorithm developed in this paper will be useful in (numerically) solving this open problem.

\subsection{Reciprocity of Alignment}
An interesting observation from the problem formulation above is the duality relationship between interference alignment on a given interference channel and its reciprocal channel obtained by switching the direction of communication. Specifically, let $\overleftarrow{\bf V}^{[k]}, \overleftarrow{\bf U}^{[k]}$ denote the transmit precoding filters and the receive interference suppression filters on the reciprocal channel. The feasibility conditions on the reciprocal channel are:
\begin{eqnarray}
\overleftarrow{\bf V}^{[k]}:N^{[k]}\times d^{[k]},&&\overleftarrow{\bf U}^{[k]\dagger}\overleftarrow{\bf U}^{[k]}={\bf I}_{d^{[k]}}\\
\overleftarrow{\bf U}^{[k]}:M^{[k]}\times d^{[k]},&&\overleftarrow{\bf U}^{[k]\dagger}\overleftarrow{\bf U}^{[k]}={\bf I}_{d^{[k]}}
\end{eqnarray}
such that
\begin{eqnarray}
\overleftarrow{\bf U}^{[j]\dagger}\overleftarrow{\bf H}^{[jk]}\overleftarrow{\bf V}^{[k]} &=& 0, \forall j\neq k\\
\mbox{rank}\left(\overleftarrow{\bf U}^{[k]\dagger}\overleftarrow{\bf H}^{[kk]}\overleftarrow{\bf V}^{[k]}\right)&=&d_k, ~\forall k\in\mathcal{K}
\end{eqnarray}
Suppose we set $\overleftarrow{\bf V}^{[k]}={\bf U}^{[k]}, \overleftarrow{\bf U}^{[k]}={\bf V}^{[k]}$. Then the feasibility conditions on the reciprocal channel become identical to the original feasibility conditions. Thus, the following observation can be made:

{\bf Reciprocity of Alignment:} {\it Since the feasibility conditions are identical, if the degrees of freedom allocation $(d^{[1]}, d^{[2]}, \cdots, d^{[K]})$ is feasible on the original interference network then it is also feasible on the reciprocal network (and vice versa). Interference alignment on the reciprocal interference network is simply achieved by choosing the transmit filters and the receive filters on the reciprocal channel as the receive filters and the transmit filters (respectively) of the original channel.}

Reciprocity of alignment is a key property used for distributed interference alignment algorithms, described in the next section.

\section{Distributed algorithm for interference alignment}
\label{sec: algo}
In this section we construct distributed interference alignment algorithms for the interference channel with multiple-antenna nodes and no symbol extensions.
\begin{figure*}
\center
\input{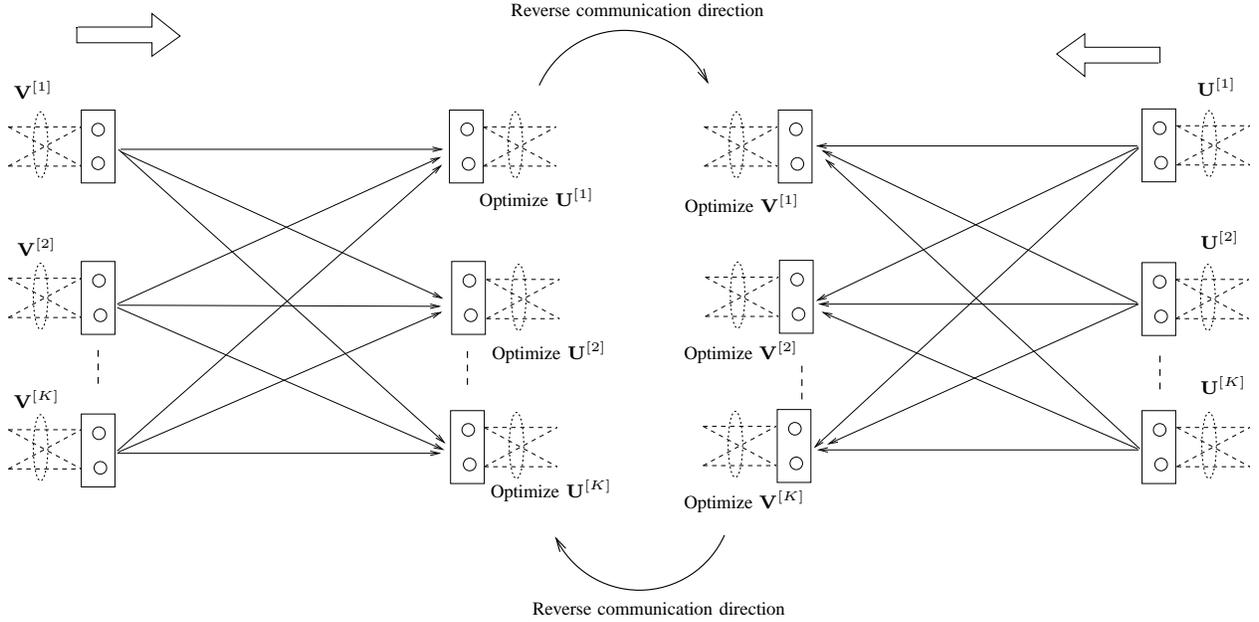}
\caption{Pictorial representation of the iterative interference
alignment algorithm where the receive directions are
optimized to minimize interference power at the receivers. Each link
(arrow) represents a MIMO channel. The transmit power per node is
$P$ in both directions. }\label{fig: algo}
\end{figure*}
Continuing with the system model of Section \ref{sec:recalign}, this implies that the relevant interference alignment feasibility condition is (\ref{eq:cross}) while (\ref{eq:direct}) is automatically satisfied. Basically, (\ref{eq:cross}) requires that at each receiver, all interference is suppressed, leaving as many interference-free dimensions as the degrees of freedom allocated to that receiver. 

Since we are interested in distributed algorithms, we start with arbitrary transmit and receive filters ${\bf V}^{[k]}, {\bf U}^{[k]}$ and iteratively update these filters to approach interference alignment. The quality of alignment is measured by the power in the \emph{leakage interference} at each receiver, i.e. the interference power remaining in the received signal after the receive interference suppression filter is applied. The goal is to achieve interference alignment by progressively reducing the leakage interference. If interference alignment is feasible then eventually leakage interference will be zero. 

The total interference leakage  at receiver $k$ due to all undesired transmitters ($j \neq k$) is  given by:
\begin{eqnarray}
I^{[k\star]}=\mbox{Tr}\left[{\bf U}^{[k]\dagger}{\bf Q}^{[k]}{\bf U}^{[k]}\right]
\end{eqnarray}
where
\begin{eqnarray}
{\bf Q}^{[k]}=\sum_{j=1,j\neq k}^K \frac{P^{[j]}}{d^{[j]}}{\bf H}^{[kj]}{\bf V}^{[j]}{\bf V}^{[j]\dagger}{\bf H}^{[kj]\dagger}
\end{eqnarray}
is the interference covariance matrix at receiver $k$.

Similarly, in the reciprocal network, the total interference leakage  at receiver $j$ due to all undesired transmitters ($k \neq j$) is  given by:
\begin{eqnarray}
\overleftarrow{I}^{[j\star]}=\mbox{Tr}\left[\overleftarrow{\bf U}^{[j]\dagger}\overleftarrow{\bf Q}^{[j]}\overleftarrow{\bf U}^{[j]}\right]
\end{eqnarray}
where
\begin{eqnarray}
\overleftarrow{\bf Q}^{[j]}=\sum_{k=1,k\neq j}^K \frac{\overleftarrow{P}^{[k]}}{d^{[k]}}\overleftarrow{\bf H}^{[jk]}\overleftarrow{\bf V}^{[k]}\overleftarrow{\bf V}^{[k]\dagger}\overleftarrow{\bf H}^{[jk]\dagger}
\end{eqnarray}
is the interference covariance matrix at receiver $j$.

The iterative algorithm alternates between the original and reciprocal networks. Within each network only the receivers update their interference suppression filters to minimize their total leakage interference. 

{\it Step I:} In the original network, each receiver solves the following optimization problem.
\begin{eqnarray}
\min_{{\bf U}^{[k]}:N^{[k]}\times d^{[k]}, ~~{\bf U}^{[k]}{\bf U}^{[k]\dagger}={\bf I}_{d^{[k]}}} I^{[k\star]}
\end{eqnarray}
In other words, receiver $k$ chooses its interference suppression filter ${\bf U}^{[k]}$ to minimize the leakage interference due to all undesired transmitters. The $d^{[k]}$ dimensional received signal subspace that contains the least interference is the space spanned by the eigenvectors corresponding to the $d^{[k]}$ smallest eigenvalues of the interference covariance matrix ${\bf Q}^{[k]}$. Thus, the $d^{[k]}$ columns of ${\bf U}^{[k]}$ are given by:
\begin{equation}{\bf U}^{[k]}_{\star d}={\bf \nu}_{d}[{\bf Q}^{[k]}],~~~~ d=1, \cdots,d^{[k]} \label{eqn: basis_vectors}\end{equation}
where ${\bf \nu}_{d}[{\bf A}]$ is the eigenvector corresponding to the $d^{th}$ smallest eigenvalue of $\bf A$.

{\it Step II:} The second step is identical to the first step, but performed in the reciprocal network. Consider the reciprocal network obtained by reversing the roles of the transmitters and the receivers. The transmit precoding matrices in the reciprocal network, $\overleftarrow{\bf V}^{[k]}$,  are the receive interference suppression matrices ${\bf U}^{[k]}$ from the original network that were determined in Step I. Each receiver in the reciprocal network solves the following optimization problem.
\begin{eqnarray}
\min_{\overleftarrow{\bf U}^{[j]}:M^{[j]}\times d^{[j]}, ~~\overleftarrow{\bf U}^{[j]}\overleftarrow{\bf U}^{[j]\dagger}={\bf I}_{d^{[j]}}} \overleftarrow{I}^{[j\star]}
\end{eqnarray}
Similar to Step I, the $d^{[j]}$ columns of $\overleftarrow{\bf U}^{[j]}$ are given by:
\begin{equation}\overleftarrow{\bf U}^{[j]}_{\star d}={\bf \nu}_{d}[\overleftarrow{\bf Q}^{[j]}],~~~~ d=1, \cdots,d^{[j]} \end{equation}
The receive interference suppression filters in the reciprocal network are then used as the transmit precoding matrices in the original network, and the algorithm returns to Step I. The iterations continue in this manner until the algorithm converges.

The iterative procedure is summarized in Algorithm \ref{alg:iter_ia}. A pictorial representation is shown in Fig. \ref{fig: algo}.
\begin{algorithm}
\caption{Iterative interference alignment}
\label{alg:iter_ia}
\begin{algorithmic}[1]%
\State Start with arbitrary precoding matrices ${\bf V}^{[j]}: M^{[j]}\times d^{[j]}, {\bf V}^{[j]}{\bf V}^{[j]\dagger}={\bf I}_{d^{[j]}}$.
\State Begin iteration 
\State Compute interference covariance matrix at the receivers: 
\[{\bf Q}^{[k]}=\sum_{j=1,j\neq k}^K \frac{P^{[j]}}{d^{[j]}}{\bf H}^{[kj]}{\bf V}^{[j]}{\bf V}^{[j]\dagger}{\bf H}^{[kj]\dagger}
\]
\State Compute the interference suppression matrix at each receiver:
\begin{equation}{\bf U}^{[k]}_{\star d}={\bf \nu}_{d}[{\bf Q}^{[k]}],~~~~ d=1, \cdots,d^{[k]}\nonumber\end{equation}
\State Reverse the communication direction and set $\overleftarrow{\bf V}^{[k]}={\bf U}^{[k]}$.
\State Compute interference covariance matrix at the new receivers:
\[
\overleftarrow{\bf Q}^{[j]}=\sum_{k=1,k\neq j}^K \frac{\overleftarrow{P}^{[k]}}{d^{[k]}}\overleftarrow{\bf H}^{[jk]}\overleftarrow{\bf V}^{[k]}\overleftarrow{\bf V}^{[k]\dagger}\overleftarrow{\bf H}^{[jk]\dagger}
\]
\State  Compute the interference suppression matrix at each receiver:
\begin{equation}
\overleftarrow{\bf U}^{[j]}_{\star d}={\bf \nu}_{d}[\overleftarrow{\bf Q}^{[j]}],~~~~ d=1, \cdots,d^{[k]}\nonumber\end{equation}
\State Reverse the communication direction and set ${\bf V}^{[k]}=\overleftarrow{\bf U}^{[k]}$.
\State Continue till convergence.
\end{algorithmic}
\end{algorithm}

\subsection{Proof of Convergence}
We now show that the algorithm must converge. The proof also highlights the intuition behind the algorithm. 

We define a  metric called the weighted leakage interference (WLI) as:
\begin{eqnarray*}
I_{w}&=&\sum_{k=1}^{K}\sum_{j=1, j\neq k}^K\frac{\overleftarrow{P}^{[k]}}{d^{[k]}}I^{[kj]}\\
&=&\sum_{k=1}^K\sum_{j=1, j\neq k}^K\frac{\overleftarrow{P}^{[k]}}{d^{[k]}}\frac{{P}^{[j]}}{d^{[j]}}\mbox{Tr}\left[{\bf U}^{[k]\dagger}{\bf H}^{[kj]}{\bf V}^{[j]}{\bf V}^{[j]\dagger}{\bf H}^{[kj]\dagger}{\bf U}^{[k]}\right]
\end{eqnarray*}
We show that each step in the algorithm reduces the value of WLI. Since WLI is bounded below by zero, this implies that the algorithm must converge. Note that an interference alignment solution corresponds to WLI$=0$. 

The WLI associated with receiver $k$ is
\begin{eqnarray}
I^{[k\star]}_{w}&=&\frac{\overleftarrow{P}^{[k]}}{d^{[k]}}\sum_{j=1, j\neq k}^K\frac{{P}^{[j]}}{d^{[j]}}\mbox{Tr}\left[{\bf U}^{[k]\dagger}{\bf H}^{[kj]}{\bf V}^{[j]}{\bf V}^{[j]\dagger}{\bf H}^{[kj]\dagger}{\bf U}^{[k]}\right]\nonumber  \label{eqn:leakkstar}\\
&=&\frac{\overleftarrow{P}^{[k]}}{d^{[k]}}\mbox{Tr}\left[{\bf U}^{[k]\dagger}{\bf Q}^{[k]}{\bf U}^{[k]}\right]=\frac{\overleftarrow{P}^{[k]}}{d^{[k]}}I^{[k\star]}\nonumber  
\end{eqnarray}
Therefore the value of ${\bf U}^{[k]}$ computed in Step 4 to minimize $I^{[k\star]}$ also minimizes $I^{[k\star]}_{w}$. Since $I_w=\sum_{k=1}^KI^{[k\star]}_w$, we have
\begin{eqnarray*}
\min_{{\bf U}^{[1]},{\bf U}^{[2]}, \cdots, {\bf U}^{[K]}} I_w &=& \min_{{\bf U}^{[1]},{\bf U}^{[2]}, \cdots, {\bf U}^{[K]}} \sum_{k=1}^KI_w^{[k\star]}\\
&=&\sum_{k=1}^K \left[\min_{{\bf U}^{[k]}}I_w^{[k\star]}\right]=\sum_{k=1}^K \frac{\overleftarrow{P}^{[k]}}{d^{[k]}}\left[\min_{{\bf U}^{[k]}}I^{[k\star]}\right]
\end{eqnarray*}
In other words, given the values of ${\bf V}^{[j]}, j\in\{1,2,\cdots, K\}$, Step 4 minimizes the value of $I_w$ over all possible choices of ${\bf U}^{[k]}, k\in\{1,2,\cdots, K\}$. In particular, Step 4 can only reduce the value of $I_w$.

The weighted leakage interference associated with transmitter $j$ is
{\allowdisplaybreaks
\begin{eqnarray*}
I^{[\star j]}_w&=&\frac{{P}^{[j]}}{d^{[j]}}\sum_{k=1}^K\frac{\overleftarrow{P}^{[k]}}{d^{[k]}}\mbox{Tr}\left[{\bf U}^{[k]\dagger}{\bf H}^{[kj]}{\bf V}^{[j]}{\bf V}^{[j]\dagger}{\bf H}^{[kj]\dagger}{\bf U}^{[k]}\right]\\
&=&\frac{{P}^{[j]}}{d^{[j]}}\sum_{k=1}^K\frac{\overleftarrow{P}^{[k]}}{d^{[k]}}\mbox{Tr}\left[\overleftarrow{\bf V}^{[k]\dagger}\overleftarrow{\bf H}^{[jk]\dagger}\overleftarrow{\bf U}^{[j]}\overleftarrow{\bf U}^{[j]\dagger}\overleftarrow{\bf H}^{[jk]}\overleftarrow{\bf V}^{[k]}\right]\\
&=&\frac{{P}^{[j]}}{d^{[j]}}\sum_{k=1}^K\frac{\overleftarrow{P}^{[k]}}{d^{[k]}}\mbox{Tr}\left[\overleftarrow{\bf U}^{[j]\dagger}\overleftarrow{\bf H}^{[jk]}\overleftarrow{\bf V}^{[k]}\overleftarrow{\bf V}^{[k]\dagger}\overleftarrow{\bf H}^{[jk]\dagger}\overleftarrow{\bf U}^{[j]}\right]\\
&=&\frac{{P}^{[j]}}{d^{[j]}}\mbox{Tr}\left[\overleftarrow{\bf U}^{[j]\dagger}\overleftarrow{\bf Q}^{[j]}\overleftarrow{\bf U}^{[j]}\right]\\
\end{eqnarray*}
}
Therefore the value of $\overleftarrow{\bf U}^{[j]}$ computed in Step 7  to minimize $\overleftarrow{I}^{[j\star]}$ also minimizes $I^{[\star j]}_{w}$. Since $I_w=\sum_{k=1}^KI^{[\star j]}_w$, it is easily seen that Step 7 can also only reduce the value of $I_w$. Since the value of $I_w$ is monotonically reduced after every iteration, convergence of the algorithm is guaranteed.

{\it Remark:} While the algorithm minimizes leakage interference at every iteration and is guaranteed to converge, convergence to global minimum is not guaranteed due to the non-convex nature of the interference optimization problem. Numerical results for the performance of the algorithm are presented in the next section.

The following observations summarize the intuition behind the iterative algorithm.
\begin{enumerate}
\item Dimensions along which a receiver sees the least interference from other nodes are also the dimensions along which it causes the least interference to other nodes in the reciprocal network where it functions as a transmitter.
\item The weighted leakage interference is unchanged in the original and reciprocal networks if the transmit and receive filters are switched.
\end{enumerate}

\subsection{Max-SINR Algorithm}
The algorithm presented above seeks perfect interference alignment. In particular it seeks to create an interference-free subspace of the required number of dimensions, that is designated as the \emph{desired} signal subspace. However, note that interference alignment makes no attempt to maximize the desired signal power within the desired signal subspace. In fact the algorithm described above does not depend at all on the direct channels ${\bf H}^{[kk]}$ through which the desired signal arrives at the intended receiver. Therefore, while the interference is eliminated within the desired space, no coherent combining gain (array gain) for the desired signal is obtained with interference alignment. While this is optimal as all signal powers approach infinity, it is not optimal in general at intermediate SNR values. Therefore other algorithms may be designed which will perform better than the interference alignment algorithm at intermediate SNR values.

In this section we consider one such  natural extension of the interference alignment algorithm where the receive filters ${\bf U}^{[k]}$ and $\overleftarrow{\bf U}^{[k]}$ are chosen  to maximize SINR at the receivers instead of only minimizing the leakage interference. While there is no loss of generality in assuming orthogonal precoding vectors for the streams sent from the same transmitter as far as interference alignment is concerned, orthogonal precoding vectors are in general suboptimal for SINR optimization. We therefore no longer assume that the columns of ${\bf V}^{[k]}$ (the transmit precoding vectors) are mutually orthogonal. We also identify the columns of ${\bf U}^{[k]}$ to be the specific combining vectors for the corresponding desired data stream, so that they are not necessarily orthogonal either. With these modified definitions, the SINR of the $l^{th}$ stream of the $k^{th}$ receiver is
\begin{equation}
\mbox{SINR}_{kl}=\frac{{\bf U}^{[k]\dagger}_{\star l}{\bf H}^{[kk]}{\bf V}^{[k]}_{\star l}{\bf V}^{[k]\dagger}_{\star l}{\bf H}^{[kk]\dagger}{\bf U}^{[k]}_{\star l}}{{\bf U}^{[k]\dagger}_{\star l}{\bf B}^{[kl]}{\bf U}^{[k]}_{\star l}}\frac{P^{[k]}}{d^{[k]}}
\end{equation}
where 
\begin{eqnarray}
\small {\bf B}^{[kl]}&=&
\sum_{j=1}^{K}\frac{P^{[j]}}{d^{[j]}}\sum_{d=1}^{d^{[j]}}{\bf H}^{[kj]}{\bf V}^{[j]}_{\star d}{\bf V}^{[j]\dagger}_{\star d}{\bf H}^{[kj]\dagger}\nonumber\\
&& ~~-\frac{P^{[k]}}{d^{[k]}}{\bf H}^{[kk]}{\bf V}^{[k]}_{\star l}{\bf V}^{[k]\dagger}_{\star l}{\bf H}^{[kk]\dagger}+{\bf I}_{N^{[k]}}\label{eq:Bkl}
\end{eqnarray}

The unit vector ${\bf U}^{[k]}_{\star l}$ that maximizes $\mbox{SINR}_{kl}$ is given by
\begin{equation}
{\bf U}^{[k]}_{\star l}= \frac{\left({\bf B}^{[kl]}\right)^{-1}{\bf H}^{[kk]}{\bf V}^{[k]}_{\star l}}{\parallel \left({\bf B}^{[kl]}\right)^{-1}{\bf H}^{[kk]}{\bf V}^{[k]}_{\star l}\parallel}.\label{eq:Ukl}
\end{equation}

The steps of the iteration are given in Algorithm \ref{alg:max_sinr}. 

\begin{algorithm}
\caption{Max-SINR algorithm }
\label{alg:max_sinr}
\begin{algorithmic}[1]%
\State Start with any ${\bf V}^{[k]}:M^{[k]}\times d^{[k]}$, columns of ${\bf V}^{[k]}$ are linearly independent unit vectors.

\State Begin iteration \State Compute interference plus noise covariance matrix
for ${\bf B}^{[kl]}$ for stream $l$ at receiver $k$ according to (\ref{eq:Bkl}), $\forall ~k\in\{1,2,\cdots, K\}, l\in\{1,2,\cdots,d^{[k]}\}$.

\State Calculate receive combining vectors ${\bf U}^{[k]}_{\star l}$ at receiver $k$ according to (\ref{eq:Ukl}), $\forall ~k\in\{1,2,\cdots, K\}, l\in\{1,2,\cdots,d^{[k]}\}$.

\State Reverse the communication direction and use the receive
combining vectors as precoding vectors: $\overleftarrow{\bf V}^{[k]}={\bf U}^{[k]}$, $\forall ~k\in\{1,2,\cdots, K\}$.
\State In the reciprocal network, compute interference
plus-noise covariance matrix $\overleftarrow{\bf B}^{[kl]}$ for stream $l$ at receiver $k$, $\forall ~k\in\{1,2,\cdots, K\}, l\in\{1,2,\cdots,d^{[k]}\}$.

\State Calculate receive combining vectors $\overleftarrow{\bf U}^{[k]}_{\star l}$, $\forall ~k\in\{1,2,\cdots, K\}, l\in\{1,2,\cdots,d^{[k]}\}$.
\State Reverse the communication direction and use the receive
combining vectors as precoding vectors: ${\bf V}^{[k]}=\overleftarrow{\bf U}^{[k]}$, $\forall ~k\in\{1,2,\cdots, K\}$.
\State Repeat until convergence.
\end{algorithmic}
\end{algorithm}

\section{Performance Results and Applications}
\label{sec: results}
\begin{figure}
\center
\centerline{\psfig{figure=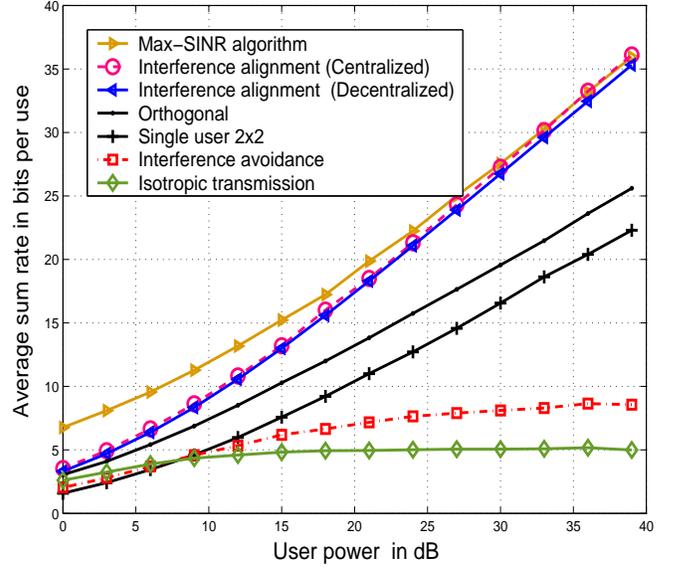,width=3.4in,height=3in}}
\caption{Performance of the decentralized interference alignment
algorithm for the three user two antenna case.} \label{algo_3u_2a}
\end{figure}

Consider the $3$ user interference channel where each node is equipped with $2$ antennas and all channel coefficients are i.i.d. zero mean unit variance circularly symmetric complex Gaussian. As shown in Fig. \ref{fig: nash} with interference alignment each user achieves $1$ degree of freedom. In Fig. \ref{algo_3u_2a} we compare the performance of the iterative interference alignment algorithm for this channel with the theoretical interference alignment which assumes global channel knowledge. It can be seen that the algorithm performs very  close to the theoretical case and more importantly it provides significant benefits over the orthogonal case. The sum rate for the orthogonal scheme is calculated assuming equal time sharing for the users, and with power $3P$ per node. The modified interference alignment scheme where the receive-combining vectors maximize the SINR provides a considerable performance gain at low and intermediate $P$. As expected, it achieves the same performance of interference alignment at high $P$.  We also plot the performance of interference avoidance algorithm  which is a selfish approach. The isotropic transmission case refers to the case where each transmitter sends $2$ streams of equal power without regard to the channel information. 
In the following, we explore some of the applications of the iterative interference alignment algorithm.
\subsection{Feasibility of Interference Alignment}
While the iterative algorithm is useful for circumventing the need for global channel knowledge, it can also be used to check theoretical feasibility of interference alignment for a given number of streams per user. Let $(d^{[1]},d^{[2]}, \cdots d^{[k]})$ denote the number of transmit streams of the users. 
For perfect interference alignment $\sum_{j=1}^{d^{[k]}}\lambda_j[{\bf  Q}^{[k]}]=0$ at receiver $k$ where $\lambda_j[{\bf A}]$ denotes the $j$th smallest eigenvalue of $\bf A$. Note that $\sum_{j=1}^{d^{[k]}}\lambda_j[{\bf Q}^{[k]}]$ indicates the interference power in the desired signal space.

\begin{figure}
\center
\centerline{\psfig{figure=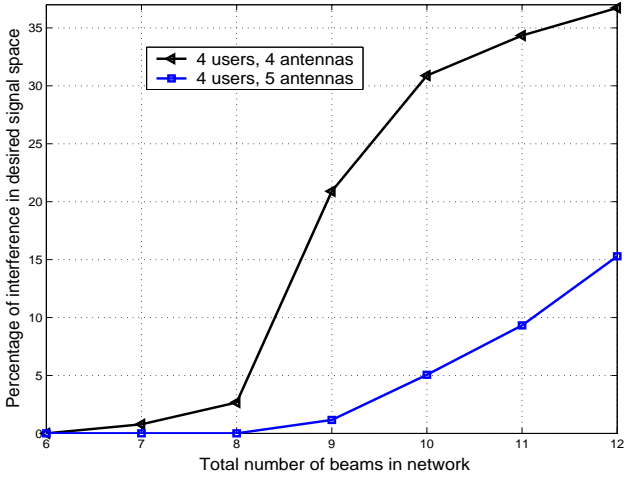,width=3.3in,height=2.5in}}
\caption{Percentage of interference power in desired signal space as
a function of the total number of data streams in the network.}
\label{int_percent}
\end{figure}

Using the algorithm, we plot in Fig. \ref{int_percent}, the percentage of interference in
the desired signal space versus the total
number of transmit streams in the network. The fraction of interference in
the desired signal space of receiver $k$ is defined as
\begin{equation}
p_{k}= \frac{\sum_{j=1}^{d^{[k]}}\lambda_j[{\bf  Q}^{[k]}]}{\mbox{Tr}[{\bf Q}^{[k]}]}.
\end{equation}
 When interference alignment is feasible the
fraction of interference in desired signal space will be zero (within numerical errors). Fig. \ref{int_percent} suggests that interference alignment
is feasible on the four user interference channel with 5 antennas at each  node when each transmitter sends two streams. With increase in the number of
streams the interference in desired signal space increases which is
an indication that interference alignment is not possible. Although
the upperbound on the degrees of freedom for this network is 10, approaching the upperbound may entail channel extensions. The plot suggests that 8 degrees of freedom ($d^{[1]}=d^{[2]}=d^{[3]}=d^{[4]}=2$) can be achieved without channel extension. Similarly for the 4 antenna case, the plot indicates that interference alignment is possible for up to a total of 6 streams in the $4$ user interference network with only $4$ antennas at each node. 

\subsection{Networks with single antenna nodes}
MIMO nodes are not necessary in order to achieve interference alignment in wireless networks. Interference can be aligned even in networks with single antenna nodes through channel extension in frequency or time as long as the channel is varying across frequency or time \cite{Cadambe_Jafar_int}. However, one caveat with this approach is the need for long symbol extensions. Aligning interference with a large number of beams over a large number of dimensions can be especially challenging for iterative algorithms due to large dimensionality of the optimization space and the inherently non-convex nature of the problem. Moreover, symbol extensions over orthogonal dimensions produce structured (diagonal or block diagonal) matrices for which both conditions (\ref{eq:direct}) and (\ref{eq:cross}) are non-trivial. Due to these difficulties, it is preferred if interference alignment can be accomplished without symbol extensions or with limited symbol extensions.

\begin{figure}
\center
\input{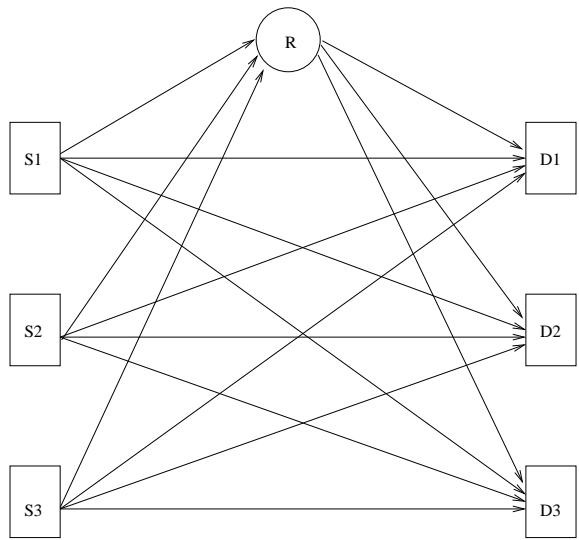}
\caption{Interference relay channel} \label{fig: relay_int}
\end{figure}

In this section, we give an example of how long symbol extensions may be avoided by the use of relays. Note that \cite{Cadambe_Jafar_XFB} has shown that relays cannot increase the degrees of freedom for time-varying wireless networks. However, as we show in this section, relays can be very useful by reducing the size of the signalling space over which interference alignment can be accomplished. The key idea is to employ relays to create a \emph{virtual} MIMO system. Consider an interference relay channel with three sources, three destinations and a half-duplex relay (node $0$) as shown in Fig. \ref{fig: relay_int}. Recall that in the absence of relays this network is shown to approach the upperbound of $3/2$ degrees of freedom per orthogonal dimension in the asymptotic limit of infinitely long symbol extensions \cite{Cadambe_Jafar_int}. However, we show that with relays
only a two time slots are required to achieve the outerbound, i.e. $3/2$ degrees of freedom. 

Consider the following two slot protocol. In the first slot, the relay is silent and the  received signal at destination $j$ is given by
\[y^{[j]}(1)=\sum_{i=1}^{3}h^{[ji]}(1)x^{[i]}(1)+z^{[j]}(1), ~~~j=1,2,3\]
The received signal at the relay can be expressed as
\[y^{[0]}(1)=\sum_{i=1}^{3}{h}^{[0i]}(1)x^{[i]}(1)+z^{[0]}(1)\] 

In the second slot, the relay transmits a scaled version of its received symbol while source $i$ transmits $x^{[i]}(2)$. The received signal at the $j^{th}$
destination node in the second slot ($j=1,2,3$)
\[y^{[j]}(2)=\sum_{i=1}^{3}{h}^{[ji]}(2)X^{[i]}(2)+{h}^{[j0]}(2)\beta y^{[0]}(1)+z^{[j]}(2)\]
Let ${Y}^{[j]}=[y^{[j]}(1)~ y^{[j]}(2)]^{T}$ and 
${X}^{[i]}=[x^{[i]}(1)~ x^{[i]}(2)]^{T}$.  In vector form, the received signal at destination $j$ can be expressed as
\begin{equation}
{Y}^{[j]}=\sum_{i=1}^{3}{\bf H}^{[ji]}{X}^{[i]}+ {Z}^{[j]}
\end{equation}
where
$$\small{\bf H}^{[ji]}=\left[\begin{array}{cc} h^{[ji]}(1) &0\\
\beta h^{[j0]}(2)h^{[0i]}(1) & h^{[ji]}(2)
\end{array}\right]$$
and
$${Z}^{[j]}=\left[\begin{array}{c} z^{[j]}(1) \\ \beta h^{j0}(2)z^{[0]}(1) +z^{[j]}(2)\end{array}\right]$$

Thus, over two time slots, the relay network reduces to a three user MIMO interference channel with (non-diagonal) structure on the channel matrix. Since the channel matrix is non-diagonal (unlike symbol extensions in the absence of a relay) it is easy to verify that when the channels are random and independent of
each other, a multiplexing gain of $\frac{3}{2}$ is achieved with probability $1$. The advantage of this scheme is that it requires only two time-slots to achieve $3/2$ degrees of freedom per time-slot, whereas in the absence of the relay, infinitely many time-slots are used in \cite{Cadambe_Jafar_int}. Since fewer dimensions are needed, iterative algorithms work better (faster convergence) in this setting. 

\begin{figure}
\center
\centerline{\psfig{figure=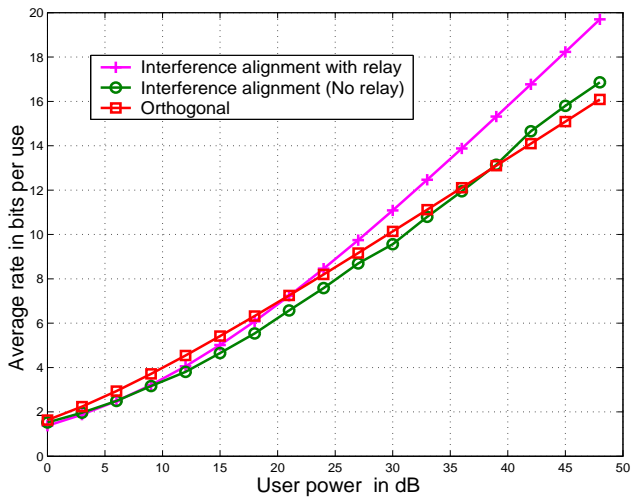,width=3.3in,height=2.6in}}
\caption{Performance of the interference alignment schemes with time slot extension for the three user interference channel.}
\label{int_relay}
\end{figure}

In Fig. \ref{int_relay} we plot the performance of the interference alignment schemes for the case of time extensions. For the interference alignment scheme with relay, the transmit power per node is $P$. That is, the three transmitters and the relay have a total power of $4P$. The transmit power for the orthogonal scheme where only transmitter is active at a time is $4P$. The interference alignment scheme without relay achieves a multiplexing gain of 4 in 3 time slots \cite{Cadambe_Jafar_int}. Here transmitter 1 sends two streams while transmitters 2 and 3 send one stream each. The transmit power per stream is $P$ adding up to $4P$ for the scheme. We can also increase the number of time slots for channel extension to improve the multiplexing gain. However it requires very high $P$ to outperform the orthogonal scheme. It can be seen that adding an extra relay helps in achieving the multiplexing gain of $\frac{3}{2}$ in two slots. Further the performance improves at low $P$ as well.

It must be stressed that the main idea behind the interference alignment scheme with the relay is to show that there are benefits in employing relays when the nodes do not have multiple antennas, especially for iterative algorithms. The scheme can be further improved by optimally allocating power to the relay and employing Algorithm \ref{alg:max_sinr} to improve its performance at low $P$.

\section{Conclusion}
\label{sec: concl}
Distributed interference alignment algorithms are investigated. Interference alignment is found to be
achievable through iterative algorithms based on  network
reciprocity and the "minimize interference to others" approach.
Numerical  comparisons to orthogonal schemes, simultaneous
transmission schemes and selfish interference avoidance schemes show
that the benefits of distributed interference alignment algorithm
are significant and close to the theoretical predictions. As mitigating interference is the fundamental problem of wireless networks, the 'do no harm' approach based algorithms have enormous applications in wireless networks.

\bibliographystyle{ieeetr}
\bibliography{biblio}
\end{document}